\documentclass[aps,prl,twocolumn, oneside]{article}   	
\usepackage[utf8]{inputenc}
\usepackage[english]{babel}
\usepackage[backend=bibtex,style=nature,doi=false,isbn=false,url=false]{biblatex}
\AtEveryBibitem{\clearfield{month}}
\AtEveryCitekey{\clearfield{month}}
\usepackage{multicol}
\usepackage[margin=0.75in]{geometry} 
\usepackage{graphicx}				
\usepackage{epstopdf}
\usepackage{amssymb}
\usepackage{amsmath}
\usepackage[table]{xcolor}
\usepackage[labelfont=bf]{caption}   	
\usepackage[scaled=0.9]{helvet}

\makeatletter
\renewcommand{\maketitle}{\bgroup\setlength{\parindent}{2pt}
\begin{flushleft}	
  \textbf{\@title}
\break
  \@author
\end{flushleft}\egroup
}
\makeatother
\begin{document}
\twocolumn[%
   
     {\huge \textbf{Trapping of Single Nano-Objects in Dynamic Temperature Fields}}\\
      \vspace{2ex}\\
       \Large \textbf{Marco Braun$^{1}$, Alois W\"urger$^{2}$, Frank Cichos$^{1}$} \\
\small \textbf{$^{1}$Molecular Nanophotonics Group, Institute of Experimental Physics I, University of Leipzig, 04103 Leipzig, Germany.}\\
\small \textbf{$^{2}$LOMA, Universit\'e de Bordeaux \& CNRS, 33405 Talence, France.}\\
\small \textbf{$^\ast$Corresponding author: cichos@physik.uni-leipzig.de}\\ 
\vspace{2ex}%
]


\noindent \textbf{In this article we explore the dynamics of a Brownian particle in a feedback-free dynamic thermophoretic trap. The trap contains a focused laser beam heating a circular gold structure locally and creating a repulsive thermal potential for a Brownian particle. In order to confine a particle the heating beam is steered along the circumference of the gold structure leading to a non-trivial motion of the particle. We theoretically find a stability condition by switching to a rotating frame, where the laser beam is at rest. Particle trajectories and stable points are calculated as a function of the laser rotations frequency and are experimentally confirmed. Additionally, the effect of Brownian motion is considered. The present study complements the dynamic thermophoretic trapping with a theoretical basis and will enhance the applicability in micro- and nanofluidic devices.}

\section{Introduction}




Single particle trapping is of high importance for long-time studies of single molecules or particles in solution without mechanical immobilization\cite{Cohen2011}. This demand led to the development of traps to counter-act Brownian motion following different approaches. Optical forces can efficiently manipulate objects with sufficiently large dielectric contrast to the solvent\cite{Ashkin1986,Grier2003,Moffitt2008}. Quadrupole traps such as the Paul trap\cite{Paul1990} have been developed over half a century to trap ions in vacuum by high-frequency electric quadrupole fields and are applied in various fields such as mass spectroscopy\cite{Douglas2005} and quantum information processing\cite{Seidelin2006}. In viscous media quadrupole traps are realized by utilizing dielectrophoresis and electrophoresis\cite{Zhang2010}. Recently, Paul trapping of single submicron-sized particles in aqueous solution has been demonstrated \cite{Guan2011,Guan2011a}. Single molecule trapping efficiency is achieved with ABEL trapping which relies on adaptively controlled electric fields\cite{Cohen2005,Fields2011,Wang2011}. Independently from the electronic properties, particles can be trapped e.g. by hydrodynamic flow \cite{Tanyeri2013} or acoustic waves \cite{Qian2013}.

Temperature gradients have also been demonstrated for particle and
macromolecular manipulation \cite{Jiang2010,Maeda2011}, since they interact on both non-ionic and charged solutes through thermophoresis, an umbrella term for thermally induced
motion at a velocity which is proportional to the temperature gradient \cite{Piazza2008,Wurger2010}.
One prominent effect which leads a charged particle going from the hot to
the cold is caused by the temperature induced perturbation of its electric
double-layer\cite{Fayolle2008}. While it is known that thermophoresis can be used to locally increase the concentration of particles or molecules\cite{Duhr2006,Duhr2006a,Duhr2006b},  recently, a method was proposed to trap a single particle in a quasi-static temperature landscape that is produced by a photothermally
heated gold structure \cite{Braun2013}. In the simplest case, such a structure consists of a
circular hole in a gold film of several microns in diameter. By illuminating this gold structure by means of an expanded laser beam, a steady-state temperature field is generated capable of trapping a single particle within a local temperature minimum in a film of solvent above the center of the circular hole. 

\begin{figure}[tbp]
\begin{center}
\includegraphics [width=0.99\columnwidth]{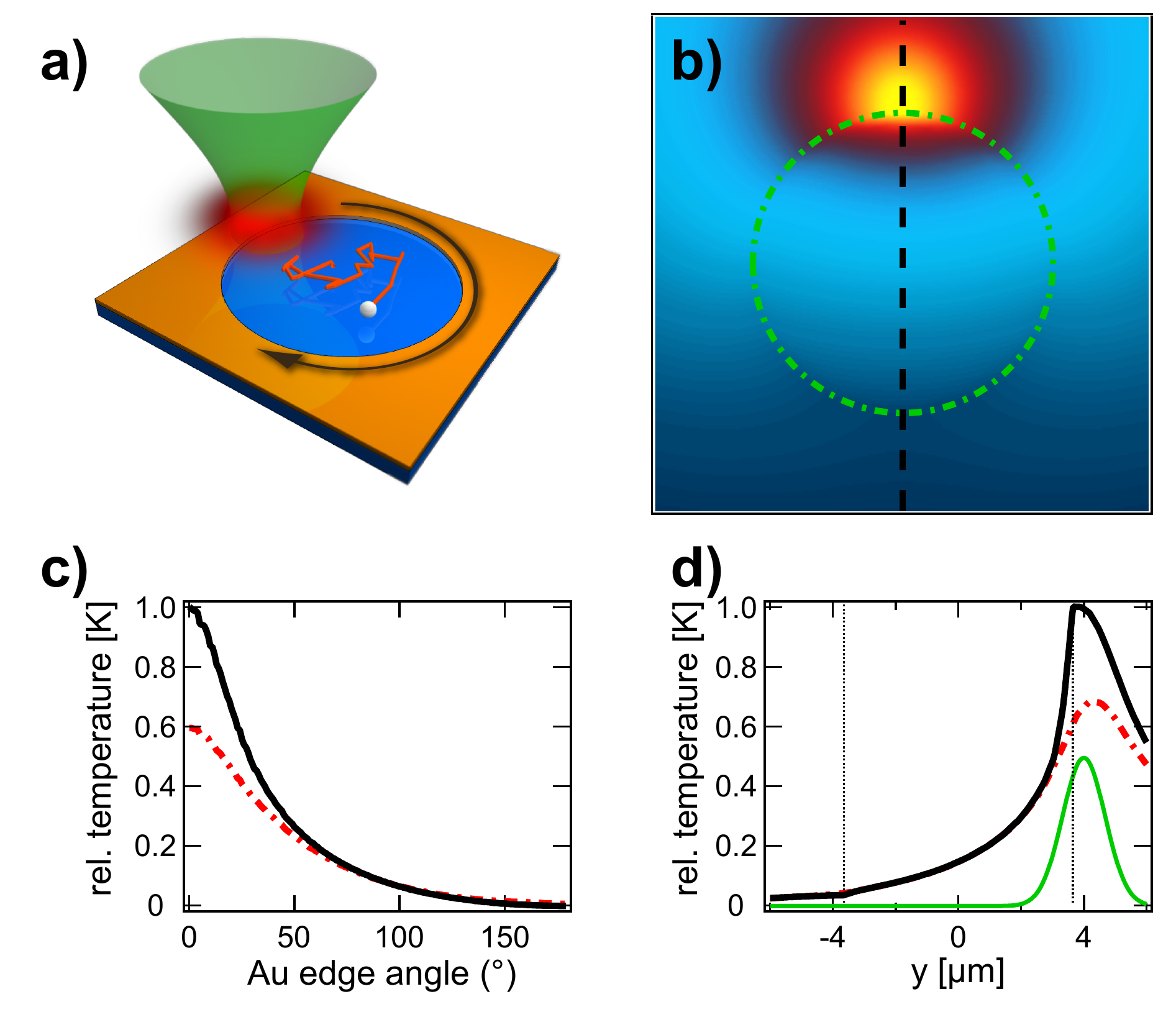}
\end{center}
\caption{\textbf{a)} Sketch of the sample: A circular gold structure is
heated by a focused laser beam. \textbf{b)} Simulated steady-state temperature field directly above the substrate
produced by the optically heated gold film by a focused laser beam. The beam is not steered, but only illuminates a local spot. The dashed green circle
indicates the edge of the gold structure (diameter $7.3\,\rm \mu m$). \textbf{c)} Temperature profile
along the gold edge in the plane of the gold structure (black) and $300\,%
\mathrm{nm}$ above (dashed red). Temperatures have been normalized to the maximum temperature induced at the gold structure. Zero temperature referes to room temperature.  \textbf{d)} temperature line profile in the
plane of the gold structure (black) and $300\,\mathrm{nm}$ above (dashed
red). The green line indicated the laser profile (in a.u.).}
\label{fig:fig1}
\end{figure}

In the present paper stronger temperature gradients are achieved by heating the edge of the Au hole using a focused laser beam, as sketched in Fig.\ \ref{fig:fig1}a. Also, for such a heating scheme, the object of interest in the center of the trap is not under direct illumination by the heating beam, preventing e.g.\ bleaching. However, for a typically positive thermodiffusive coefficient leading a particle to move to a colder region, a steady-state heating by means of a focused laser beam will end up in a purely repulsive thermal potential forcing a particle out of the trap immediately. Hence, to prevent the particle from escaping the trap, the laser beam needs to be steered. Inspired by the Paul trap and circular scanning particle trapping methods \cite{Katayama2009,Levi2003}, here, we drive the laser beam along the circumference of a hole in a gold film at a frequency $f=\omega/2\pi$ leading the thermal potential to rotate (Fig.\ \ref{fig:fig1}a). Due to a net inward component of the thermophoretic drift a confinement for a particle can be achieved in the center of the trap. 

In the following we demonstrate the feasibility of a thermophoretic particle trap using time-dependent temperature gradients. We give a detailed study of the dynamic properties of a particle. We theoretically investigate the trapping stability and determine the stationary trajectories as a function of the thermophoretic drift velocity and the rotation frequency which are experimentally confirmed. In a second step, we account for Brownian motion and determine the probability density in the thermal trapping potential. 

Thermalization of the plasmonic excitation occurs at a time-scale
of microseconds. Hence, the temperature field follows almost instantaneously the
rotation laser. Because of the different heat conductivity of gold and water,
the resulting temperature profile is not isotropic but significantly smeared
out along the edge of the gold film, as shown by the numerical simulation
results of Fig. \ref{fig:fig1}, b--d). This distortion, however, is of minor importance
for our purpose, since thermophoretic trapping relies mainly on the radial
component of the temperature gradient. Thus, the following analysis assumes
an isotropic and instantaneous profile $T(\mathbf{r,}t)=T_{0}+Q/(4\pi\kappa\left | \mathbf r-\mathbf r_{\rm L} (t) \right |),$ where $Q$ is the absorbed power, $\kappa$ the heat conductivity and $\mathbf{r}_{L}$ the position of the laser. 
Experiments were carried out in a microscopy setup using the sample preparation as presented in our previous publication \cite{Braun2013}. Further details are described in the Materials and Methods section. 

\begin{figure}[tbp]
\begin{center}
\includegraphics [width=0.99\columnwidth]{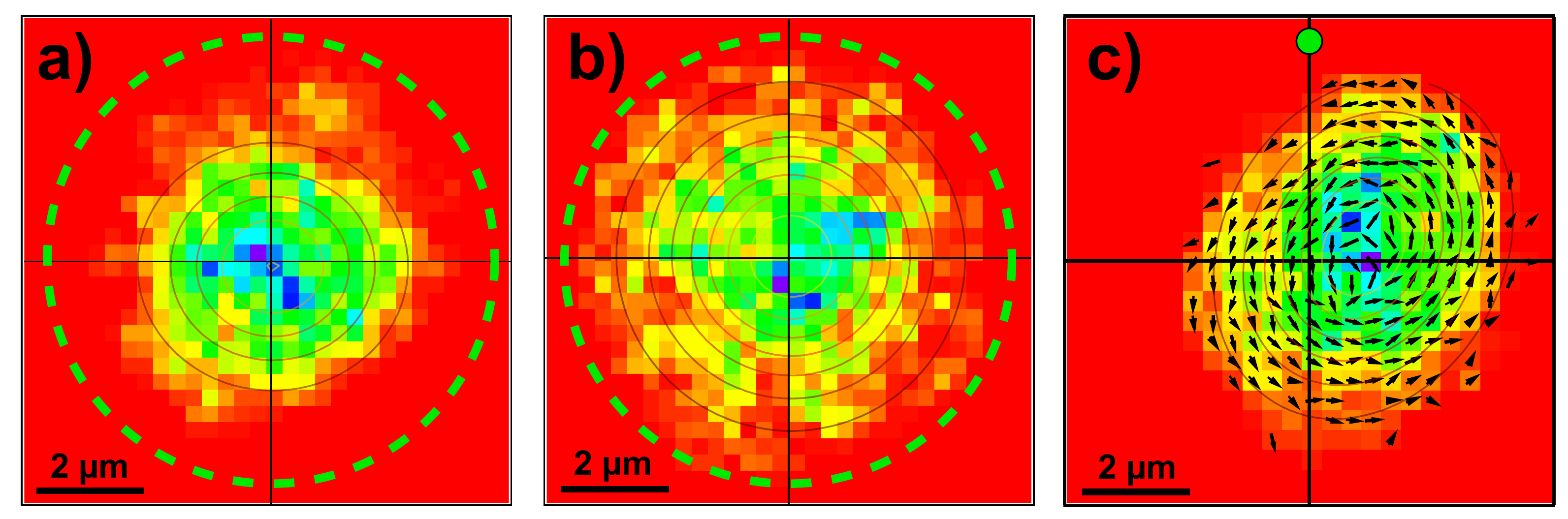}
\end{center}
\caption{\textbf{a)} Position distribution of a $460\,\mathrm{nm}$ PS sphere
within a TP trap at a laser rotation frequency of $100\,\mathrm{Hz}$. 
\textbf{b)} Same for $0.7\,\mathrm{Hz}$. The dashed green circle indicates
the path of the laser beam (in clockwise direction). \textbf{c)} Same data
as in b) after transformation to the rotating frame, where the laser is
immobile (green dot) and where the fluid rotates counter-clockwise at a
frequency of $\ 0.7\,\mathrm{Hz}$. The arrows indicate the particle velocity
field. }
\label{fig:fig2}
\end{figure}

\section{Particle Dynamics in the Rotating frame}
The experimental results of the particle dynamics in a rotating temperature field reveal some general features, which we want to highlight before starting with an in depth description of the particle motion.

Fig.\ \ref{fig:fig2}a) and b) display the positional distribution of a single $460\,\rm nm$ PS bead in the same trap structure with equal heating laser intensity but at different laser rotation frequencies of $100\,\mathrm{Hz}$ and $0.7\,\mathrm{Hz}$. Both trajectory point distributions indicate the confinement of the particle in the rotating temperature field, but the distribution for the higher frequency is narrower. 
However, while the magnitude of the thermal drift is only dependent on the heating laser intensity and does not change with rotation frequency, the inward component of the thermal drift seems to decrease for a slower rotation frequency. Due to the rotating laser field, a tangential component of the particle drift should appear as well. The particle dynamics should therefore depend on this tangential drift at slow frequency too. 
This importance of tangential and radial drift in the trap structure is better recognized when transforming the coordinate system into the frame moving with the laser beam. In such a frame the laser beam and the temperature profile are at rest but the sample including the fluid rotates counter-clockwise around the center of the trap. Fig.\ \ref{fig:fig2}c) shows the data at $0.7\,\mathrm{Hz}$ transformed to the rotating frame. The position of the heating beam is indicated by the green dot. The particles position
distribution is Gaussian but asymmetric and the maximum shifted from the
center of the trap. These features are readily understood in terms of the thermophoretic
repulsion from the laser position and the advection by the rotating flow,
and imply in particular that the particle is always in front of the laser spot.
Transforming back to the lab frame smears out the asymmetry, and one
recovers the broadened position distribution of Fig.\ \ref{fig:fig2}b). The arrows in Fig.\ \ref{fig:fig2}c) indicate the particle velocity with respect to the rotating frame. They reveal a circular motion around the
center of the distribution function. These effects will be studied in detail.

\section{Stationary points in the flow field}

The particle dynamics originates from the thermal forces, advection, and
Brownian motion. As a first step, we discard Brownian motion and retain the
deterministic part only. Due to the aqueous solvent and small particle size the Reynolds number is low $\mathrm{Re}\sim 10^{-6}$, i.e. viscous forces dominate the motion of the particle. In this over-damped limit, inertia may be neglected and the particle instantaneously follows the thermal and advection drift. 

Then the particle velocity field in the rotation frame can be written as the sum
\begin{equation}
\mathbf{u}=\mathbf v_{\rm T}+\mathbf{\omega }\times 
\mathbf{r}  \label{7}
\end{equation}%
of the thermophoretic drift velocity $\mathbf v_{\rm T}=-D_{\mathrm{T}}\nabla T(\mathbf{r})$ with the thermodiffusion coefficient $D_{\rm T}$ at the position $\mathbf{r}$ with respect to the center of the trap
and the advection by the rotating fluid with $\mathbf{\omega }=\omega \mathbf{e}_{z}$. In the following we assume an isotropic temperature
profile as mentioned above, which implies that the gradient decays as  $\mathbf{\nabla }%
T=-Q/(4\pi \kappa R^2)$, with $R=|\mathbf{r}-a\mathbf{e}_{y}|$ being the distance from the laser
position. 

\begin{figure}[tbp]
\begin{center}
\includegraphics [width=0.99\columnwidth]{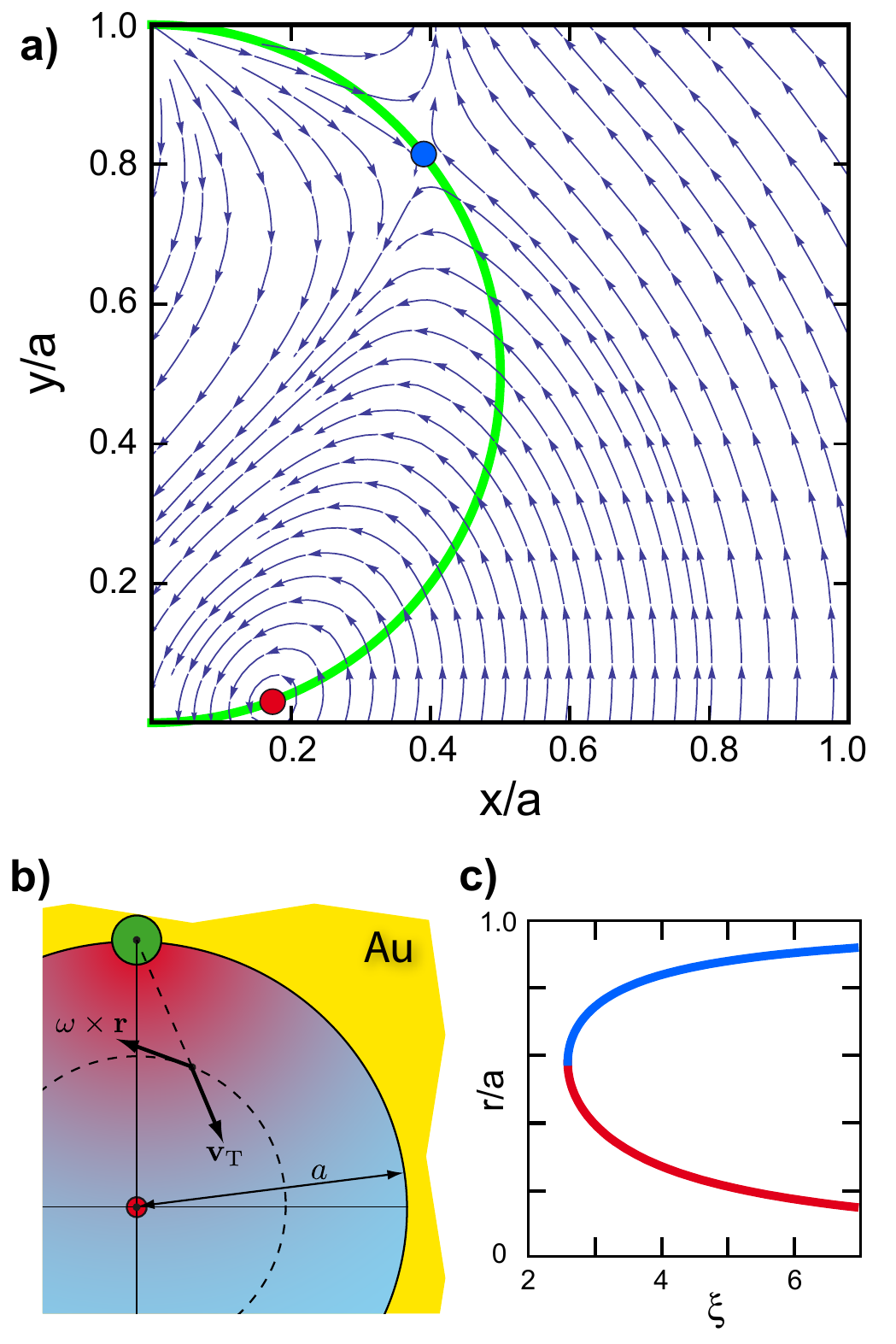}
\end{center}
\caption{\textbf{a)} Flow field in the
rotating frame $\mathbf{u}(x,y)$ for the angular frequency parameter $%
\protect\xi =6$. The green half circle indicates the line $x^{2}=y(a-y)$.
The two stationary points $\mathbf{u}=0$ are clearly visible; the lower one (red)
is an attracting stationary point, whereas the upper one (blue) is unstable. With
increasing $\protect\xi $ the stationary points repel each other; the
unstable migrates toward the position of the laser, and the stable one
toward the center. \textbf{b)} Sketch of the coordinate system used in the rotation
frame. The green dot indicates the position of the laser beam.  \textbf{c)} The curves describe the radial distance $r$ of the stationary points to the center of the gold structure as a function of the dimensionless parameter $\protect\xi$. The upper branch (blue) corresponds to an unstable stationary point, whereas the lower one (red) is stable.}
\label{fig:fig3}
\end{figure}

In Fig.\ \ref{fig:fig3}a we plot the calculated flow field $\mathbf{u}(x,y)$ in the upper-right
quadrant of the trap for a given set of parameters. The flow field shows two stationary points, where the thermophoretic drift $\mathbf v_{\rm T}$ and the advection drift $\mathbf{\omega }\times\mathbf{r}$ cancel each other such that the particle flow vanishes $\mathbf u=0$ (see Fig.\ \ref{fig:fig3}b). The upper one is unstable. A slight perturbation is amplified and the particle either escapes to infinity or moves towards the lower stationary point, which is stable. The flow field around this stationary point appears to be spiraling towards this stationary point. As the flow field is depicted in the rotating frame, a particle in this stationary point would carry out a circular motion around the trap center in the lab frame when neglecting Brownian motion. 


We further determine the position of the stationary points in the trap as a function of laser rotation frequency $\omega$ and the thermophoretic velocity ${\rm v}_{\rm T}$. Inserting the temperature gradient into equation 1 and setting u = 0 yields to the cubic equation $\xi^2 y (a-y)^2 = a^3$ and that for the half-circle, $x^2 = y(a-y)$, indicated by the green line in Fig. 3b. Their real solutions provide the positions of the above described stationary points, in terms of the dimensionless parameter
\begin{equation}
\xi =\frac{\omega a}{u_{\text{T}}},  \label{8}
\end{equation}%
which is given by the ratio of the tangential laser velocity $\omega a$ and the thermophoretic velocity 
\begin{equation*}
u_{\mathrm{T}}=D_{\mathrm{T}}\frac{Q}{4\pi\kappa a^{2}}.
\end{equation*}
The cartesian coordinates of the stable stationary point are then given by a power series in $\xi$ by $%
y_{0}=a(\xi ^{-2}+2\xi ^{-4}+...)$ and $x_{0}=\sqrt{y_{0}(a-y_{0})}$. 

Similarly, the position of the stable stationary point may also be expressed in polar coordinates given by the distance $r_{0}$ from the trap center
\begin{align}
\frac{r_{0}}{a}& =\frac{1}{\xi }+\frac{1}{\xi ^{3}}+... \label{eq:R0}
\end{align}
and the angle $\varphi_0$ when $\tan \varphi _{0}=\frac{1}{\xi }+\frac{3}{2\xi ^{3}}+... $.


The above equations immediately reveal that both stationary points exist only for a sufficiently large value of $%
\xi >\xi _{\rm min }=\sqrt{27/4}\approx 2.598$. This means that the tangential velocity of the laser on the circumference of the trap has to be larger than the thermophoretic velocity by a factor of 2.598. If this stability condition is not fulfilled, the rotating laser is to slow to prevent the particle from being pushed out of the trap by the thermal drift. In the case $\xi=\xi_{\rm min}$, both stationary points are located at the same position on the half circle. When increasing $\xi$ further they repel each other and the stable stationary point is approaching the center of the trap. With typical experimental parameters, $u_{\rm T}\sim\rm \mu m/s$ and $a\sim\rm \mu m$, one finds a minimum frequency of $\omega/2\pi\sim\rm Hz$.

\begin{figure}[tbp]
\begin{center}
\includegraphics [width=0.99\columnwidth]{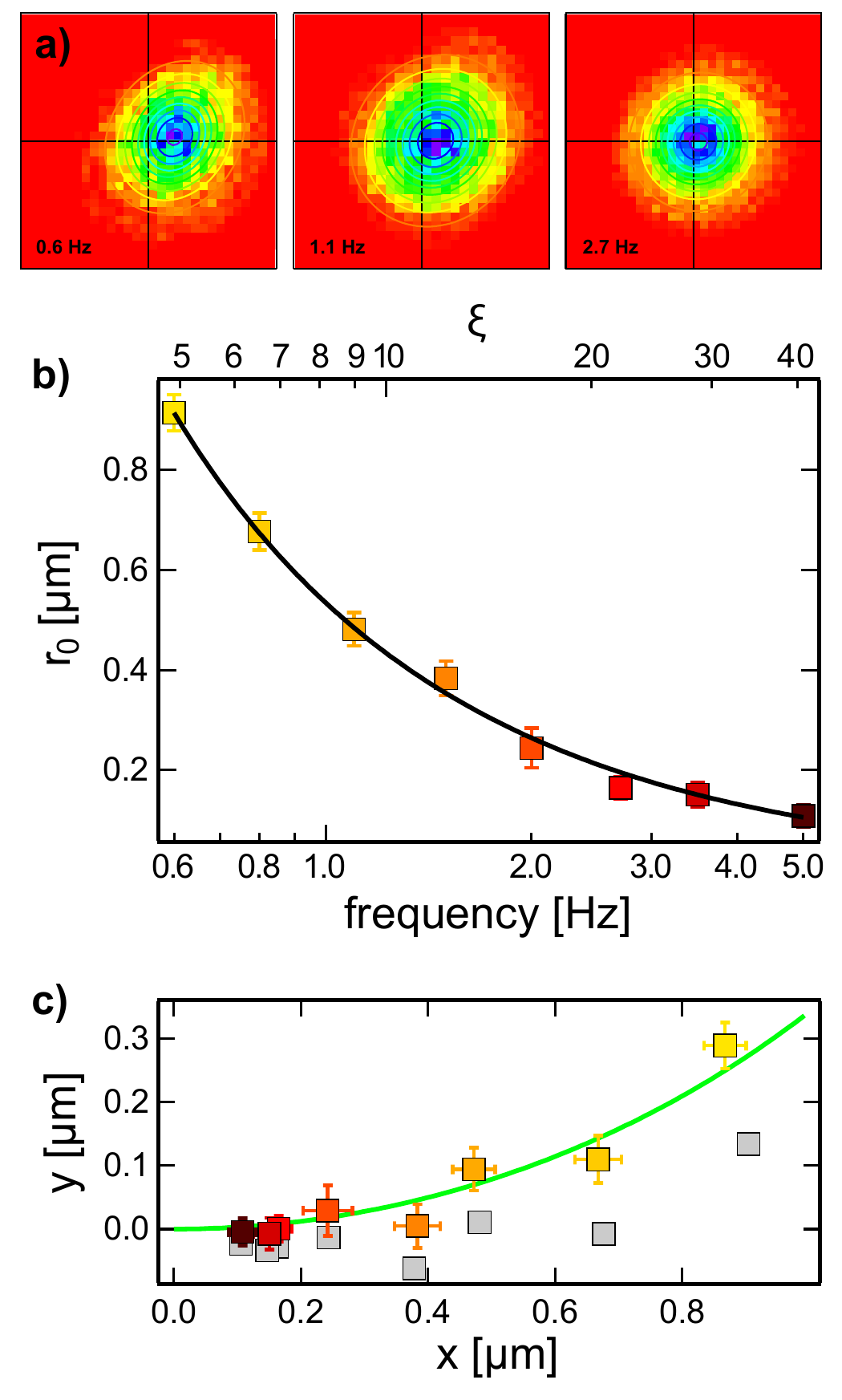}
\end{center}
\caption{\textbf{a)} Position distribution histograms in the rotating frame
for $0.6\,\mathrm{Hz}$ (left), $1.1\,\mathrm{Hz}$ (center) and $2.7\,\mathrm{%
Hz}$ (right). Each trajectory measured consists out of 10000 trajectory points. \textbf{b)} Distance of the central point of the
positional distribution as a function of the rotation frequency with fit of
equation \protect\ref{eq:R0}. \textbf{c)} Position of the central point of the positional
distributions, i.e. the positions of the stable point for different rotation
frequencies. All points are tilted by $10^\circ$ around the origin to
compensate for the finite extend of the heat source in the experiment. The
green arc again indicates the line $x^{2}=y(a-y)$. The frequency is
color-coded and can be read from the plot in c. The grey squares is the
uncorrected data.}
\label{fig:fig5}
\end{figure}

These theoretical findings agree well with experimental data obtained for a single $460 \,\mathrm{nm}$ PS bead in water recorded at different laser rotation frequencies $\omega$. At each frequency, the particle positions have been recorded and were  transformed to the rotating frame. Figure \ref{fig:fig5}a displays the corresponding histograms of the particle positions for three different laser rotation frequencies and already reveals the shift of the stationary point towards the trap center with increasing rotation frequency $\omega$. The distance of the histogram maximum for different laser rotation frequencies follows nicely the predicted frequency dependence. Fitting the radial distance as a function of frequency (eq. \ref{eq:R0}) in Fig. \ref{fig:fig5}b directly yields a thermophoretic velocity of $u_{\mathrm{T}}=(3.3\pm 0.1)\,\mathrm{\mu m/s}$ at the trap radius of $a=4.3\,\mathrm{\mu m}$. The ${x,y}$ positions of the measured maxima are consistently below the half circle $x^{2}=y(a-y)$ (Figure \ref{fig:fig5}c, grey squares). While the radial distance $r_{0}$ is matched by the theory, the phase $\varphi_0$ is preceding the theoretical phase due to the fact that the real heat source is smeared out along the rim of the gold structure (see Fig. \ref{fig:fig1}a), while we model the behavior with a  point heat source. We estimate a resulting shift in angle to be $\Delta \varphi_0\approx 10^\circ$. The corrected data is shown with the colored squared Fig.\ \ref{fig:fig5}c and follows the half circle indicated by the green line. Additionally, via the simulated temperature profile (Fig.\ \ref{fig:fig1}) and the measured thermal velocity $u_{\rm T}$, the temperature increase in the trapping center is estimated to be about $12\,\rm K$.


\section{Flow towards the stable stationary point}

While the flow field already indicates the two different stationary points we can analyze the motion of the particle close to the 
tentatively stable stationary point in more detail. We therefore linearize the flow $\mathbf{u(r)}$ at the distance from the
stationary point, $\mathbf{\hat{r}}=\mathbf{r-r}_{0}$, and then expand in
powers of $1/\xi $ 
\begin{align}
\mathbf{u} &=\mathbf{\omega }\times \mathbf{\hat{r}}+\frac{\omega }{\xi }\left( \hat{x}\mathbf{e}_{x}-2\hat{y}\mathbf{e}_{y}\right) +...  \label{14}
\end{align}%
where we have discarded terms of $O(\xi ^{-2})$.\ The first term describes the
rotation around the stationary point with frequency $\omega $.\ 

The second term, which is by a factor $\xi$ smaller and therefore independent of $\omega$, accounts for the
radial flow with respect to the stationary point at ${\bf r}_{0}$. The flow along the $\hat{x}$-direction with velocity $\omega 
\hat{x}/\xi=u_{T}\hat{x}/a$ is oriented outward, whereas along the $\hat{y}$-direction
there is an inward flow toward the stationary point with twice the velocity $%
-2\omega \hat{y}/\xi=-2u_{\rm T} \hat{y}/a $.\ When averaging over one cycle one finds that there
is a net inward flow towards the stationary point $\mathbf{r}_{0}$, which proofs the stable nature of this stationary point.

Eq. (\ref{14}) can be integrated to the following form, 
\begin{eqnarray}  \label{eq:CloseToStablePoint}
\hat{x}(t) &=&Ae^{-\Gamma t}\cos (\Omega t-\phi ),  \notag \\
\hat{y}(t) &=&Ae^{-\Gamma t}\sin (\Omega t),
\end{eqnarray}%
a spiral trajectory, where $A$ is the initial amplitude, $\Omega=\omega\sqrt{1-\phi^{2}}$ the frequency, $\Gamma=\omega/(2\xi)=u_{\mathrm{T}}/(2a)$
a damping coefficient and $\phi=\frac{3}{2}\xi^{-1}$ the phase describing the asymmetry. Terms of $O(\xi ^{-3})$ have been neglected. Without
taking thermal fluctuations into account the particle will
converge to the stationary point on a spiral in the rotating frame for $t\rightarrow\infty$ if the
stability condition $\xi >\xi _{\min }$ is fulfilled. Once the stationary point is reached, the particle travels in circles around the center of the trap in the lab frame. $\Gamma$ can be interpreted as a relaxation rate describing how fast a particles reaches the stable point, which is independent of $\omega$. Hence, while increasing $u_{\rm T}$ and $\omega$ by the same factor does not influence the position of the stable point, it amplifies the net inward flow. The phase $\phi $\ determines the skewness of the trajectory, which reduces to a circle for $\phi =0$, for high laser rotation frequencies. Note, that neither the flow field nor the positions of the stationary points depend on the size of the trapped particle. 

\begin{figure}[tbp]
\begin{center}
\includegraphics [width=0.85\columnwidth]{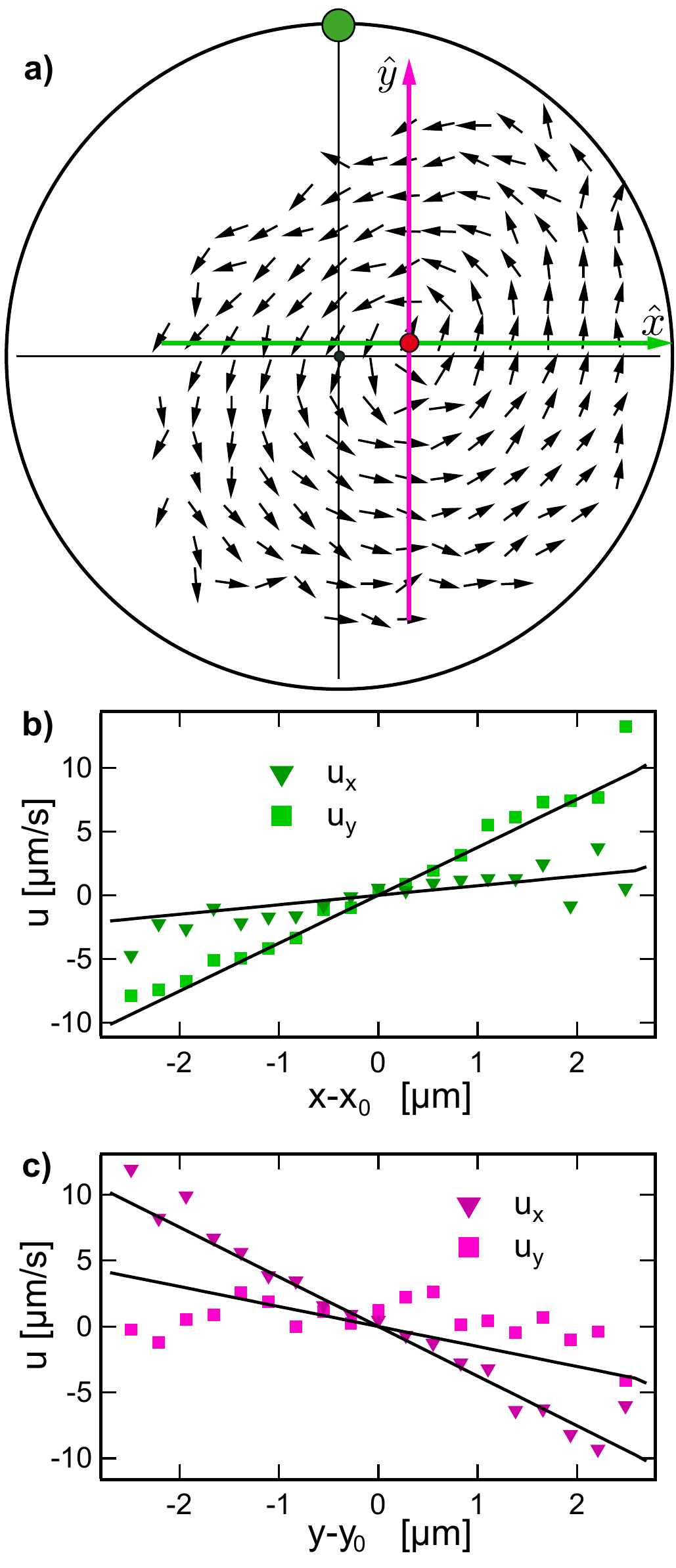}
\end{center}
\caption{\textbf{a)} Experimental flow field calculated from the trajectory for a rotation frequency of $f=0.6\,\mathrm{Hz}$. The green dot indicated the position of the heated spot. The red dot shows the location of the stationary point. Note that in this image the length of the arrow does not represent $\left | \mathbf{u }\right |$. \textbf{b)} and \textbf{c)} Flow velocities in $x$ and $y$-direction along the green and magenta lines in a). The black lines are not fits, but calculated according to eqns \protect\ref{eq:CloseToStablePoint} for a trap radius  $a=4.3\,\mathrm{\protect\mu m}$ and $u_{\mathrm{T}}=3.3\,\mathrm{\protect\mu m/s}$ as obtained earlier.}
\label{fig:fig7}
\end{figure}

A vector plot of an experimentally observed velocity field $\mathbf{u}%
(x,y)=(u_x(x,y),u_y(x,y))$ in the rotating frame is shown in Figure \ref%
{fig:fig7} for the lowest measured frequency of $f=0.6\,\mathrm{Hz}$ ($%
\xi=4.9$). Each arrow represents the average direction of the particle in
the according region, such that the stochastic Brownian motion of the
particle averages out. To compare the data to the
theoretical description, we plotted the velocities separated in $x$ and $y$%
-direction along the horizontal (green) and vertical (magenta) lines in
figures \ref{fig:fig7}b and \ref{fig:fig7}c. Correspondingly, the black
lines were calculated from equations \ref{eq:CloseToStablePoint} with $%
f=\omega/2\pi=0.6\,\mathrm{Hz}$, $a=4.3\,\mathrm{\mu m}$ and $u_{\mathrm{T}}=3.3\,\mathrm{\mu m/s}$ which was found from the fit of eqn.\ \ref{eq:R0} in Fig.\ \ref{fig:fig5}c. As can be seen, the theory and experimental data agree very well.

Although working at much lower frequencies, the motion that is observed for a particle in a thermal trap with a rotating temperature field exhibits strong similarities to the motion of ions in a Paul trap, which travel on non-trivial trajectories within the trap. Depending on the stability parameters a macro-motion is observed superimposed with the micro-motion at the frequency of the rotating quadrupole field\cite{Berkeland1998,Paul1990}. In our description of the thermal trap we decoupled the micro-motion at $\omega$ by switching to the rotating frame. Within this frame, we observe a harmonic oscillation (macro-motion) at a frequency $\Omega$ which also depends on the trapping parameters. However, due to the viscous damping at low Reynolds number in the thermal trap this macro-motion disappears exponentially and the particle reaches the stable point in the long time limit\cite{Park2012} whereas it sustains for ion trapped in vacuum. 

Eqns.\ (\ref{eq:CloseToStablePoint}) resemble a solution of a two-dimensional damped harmonic oscillator. Hence, from this trajectory it is clear that the particle is confined in an effective anisotropic harmonic potential in the rotating frame, leading to an anisotropic Gaussian positional distribution. 


\section{Diffusion and probability distribution}

So far we have not taken into account the Brownian motion of the particle.
The corresponding convection-diffusion problem is described by the
stationary Smoluchowski equation for the particle concentration,%
\begin{equation}
\nabla \cdot \mathbf J =0, \quad\mathbf J =c \mathbf{u}- D\nabla c.
\end{equation}%
Because of the rather intricate velocity field $\mathbf{u}$ there is no
general analytical solution. In the following we derive an approximate
steady-state distribution function. 

The drift velocity (\ref{14}) is linearized in powers of $\hat{x}$ and $\hat{y}$. Its radial and angular components read to leading order in $1/\xi $,%
\begin{equation}
u_{\hat{r}}=\frac{\omega }{\xi }\frac{\hat{x}^{2}-2\hat{y}^{2}}{\hat{r}},\quad  u_{\varphi }=\omega \hat{r}.
\end{equation}%
Note that the radial drift occurs outward along the $\hat{x}$-axis and
towards the center along the $\hat{y}$-axis.\ Thus, without the angular
motion, the particle would escape within the cones $\hat{x}^{2}-2\hat{y}%
^{2}>0$. Yet since both radial drift and diffusion are slow as compared to
the angular motion, the distance $\hat{r}$ changes rather little during one
cycle.\ 

Thus, we may, in a first approximation, replace the radial velocity with its
time average $\bar{u}_{\hat{r}}$. From (\ref{eq:CloseToStablePoint}) one finds $\overline{\hat{x%
}^{2}}=\frac{1}{2}\hat{r}^{2}=\overline{\hat{y}^{2}}$, and with the
definition of $\xi $ one readily has 
\begin{equation}
\bar{u}_{\hat{r}}=-u_{T}\frac{\hat{r}}{2a}.
\end{equation}
Since $u_{T}>0$, there is an effective drift towards the stationary point.
Hence, trapping arises from the superposition of the fast angular motion and
the minus sign of the mean radial velocity $\bar{u}_{\hat{r}}$. The
stationary state is obtained requiring that the radial current $\bar{J}_{%
\hat{r}}=c\bar{u}_{\hat{r}}-Ddc/d\hat{r}$ vanishes. Solving $\bar{J}_{\hat{r}%
}=0$ results in the Gaussian probability distribution $c=c_{0}e^{-\hat{r}%
^{2}/2\sigma ^{2}}$, where the mean-square distance 
\begin{equation}
\sigma ^{2}=\frac{2Da}{u_{T}}  \label{24}
\end{equation}%
is determined by the ratio of the diffusion coefficient and the
thermophoretic velocity. 

Both from the stream lines in Fig. \ref{fig:fig3} and from the trajectories (\ref{eq:CloseToStablePoint}), it
is clear, however, that $c(\hat{x},\hat{y})$ is not isotropic in the $\hat{x}%
-\hat{y}$-plane. The anisotropy is best expressed in terms of the non-zero
correlation $\overline{\hat{x}\hat{y}}=\frac{1}{2}\hat{r}^{2}\sin \phi $,
which follows directly from (\ref{eq:CloseToStablePoint}). The correlation matrix is
diagonalized by adopting skew coordinates $\hat{r}_{\pm }=(\hat{x}\pm \hat{y}%
)/\sqrt{2}$, resulting in the steady-state distribution  
\begin{equation}
c(\hat{x},\hat{y})=c_{0}\exp \left( -\frac{\hat{r}_{+}^{2}}{2\sigma _{+}^{2}}%
-\frac{\hat{r}_{-}^{2}}{2\sigma _{-}^{2}}\right) ,\label{eq:2DGaussian}
\end{equation}%
with mean-square displacements 
\begin{equation}
\sigma _{\pm }^{2}=\left( 1\pm \sin \phi \right) \frac{2Da}{u_{T}}.\label{eq:width}
\end{equation}%

By expanding in inverse powers of $\xi $, we find $\sin \phi =\frac{3}{2\xi }$.
This parameter is largest at small frequency and decreases with increasing $%
\omega $. At large frequency the widths $\sigma _{\pm }$ become equal, the
trajectory in the trap approaches a circle, and the probability distribution
reduces to (\ref{24}). While the flow field and the positions of the stationary points are indpendent of the particle size, the probability distribution width are affected by the size via the diffusion coefficient.

\begin{figure}[tbp]
\begin{center}
\includegraphics [width=0.85\columnwidth]{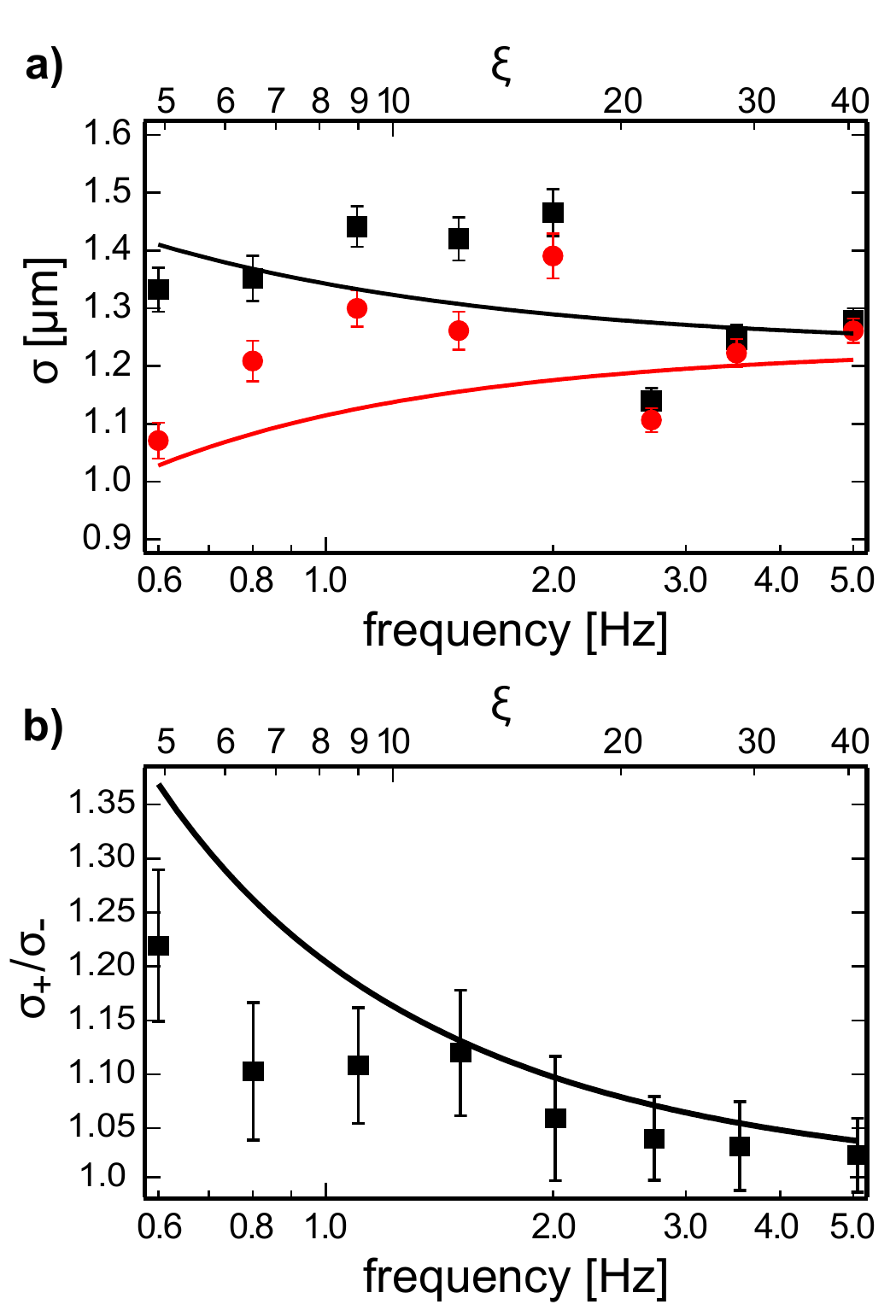}
\end{center}
\caption{\textbf{a)} Width $\protect\sigma_+$ (black) and $\protect\sigma_-$
(red) as a function of the laser rotation frequency. The curves show the
theoretical dependence of equations \protect\ref{eq:2DGaussian} and \protect
\ref{eq:width} again with $u_{\mathrm{T}}=3.3\,\mathrm{\protect\mu m/s}$ and 
$a=4.3\,\mathrm{\protect\mu m}$. \textbf{b)} Anisotropy $\protect\sigma_+/%
\protect\sigma_-$.}
\label{fig:fig8}
\end{figure}

Equations \ref{eq:2DGaussian} and \ref{eq:width} can be directly compared to
the experimental data (Figure \ref{fig:fig8}). Although the data points of $%
\sigma_+$ and $\sigma_-$ do not quantitatively follow the predictions in
Figure \ref{fig:fig8}a, it can clearly be seen that the average values of
the width are consistent with the theory. Also, the anisotropy $%
\sigma_+/\sigma_-$ is clearly visible for low rotation frequencies and
disappears for higher frequencies as expected. The main discrepancy between theory and experimental data is again attibuted to the spatially extended heat source and strong thermal conductivity of the gold layer, which disturbs the temperature profile in the experiment as compared to the modelled system, where a pure one-over-distance temperature field was assumed. 

The parameters $\sigma _{\pm }$ give the width of the trapping potential in the rotating frame. They are determined by the ratio of advective and diffusive transport rates and hence are inversely proportional to the square-root of the P\'{e}clet number $\mathrm{Pe}=u_{T}a/D$. In the experiment, with a diffusion coefficient of $D= 0.59\,\mathrm{\mu m^2/s}$, a thermal drift of $u_{\mathrm{T}}=3.3\;\mathrm{\mu m/s}$ and a trap radius of $a=4.3\,\mathrm{\mu m}$ a P\'{e}clet number of $\mathrm{Pe}\approx 24$ is achieved. The widths are also inversely proportional to the  the square-root of the Soret coefficient and excess temperature $\sigma _{\pm }\propto(S_{\mathrm T}\Delta T)^{1/2}$ similar as found in \cite{Braun2013}.



\section{Conclusion}

We have studied the motion of a single colloidal particle in a dynamic feedback-free thermal trap using a rotating temperature field to create confinement. Since the temperature field is repulsive for the colloidal particles, the confinement is the result of the dynamics of the temperature field and requires a certain threshold rotation frequency. For frequencies below this threshold particles are pushed out of the trap, while above the threshold a metastable and a stable trapping point exist. The motion of the particles around the stable stationary point is reminiscent of the complex motion in an electrodynamic Paul trap. The particle motion, however, is strongly damped as compared to the ion motion in the Paul trap due to the viscous environment. The theoretical findings are well supported by experiments confirming the main characteristics of the motion and provide a first glimpse on how single particle or even single molecule motion might be manipulated with dynamic temperature fields.

\section{Materials and Methods}

The preparation of the gold structure is fully analogous to a previous
publication \cite{Braun2013}. A clean glass substrate is coated by $5\,\rm nm$ chromium film as an adhesion layer for the gold structure. Isolated polystyrene beads ($\sim 8\,\rm\mu m$ diameter) are prepared on a glass substrate by spin coating. After coating the glass and the beads with a $50\,\rm nm$ gold layer by thermal evaporation, the beads are removed by sonication and toluene. The gold film with circular holes of about $8\,\rm\mu m$ diameter remains on the glass substrate. The chromium film uncovered by the gold is removed by etching. The experimental sample consists of two parallel glass slides, where
the lower one carries the gold structure. A water film of about $700\,%
\mathrm{nm}$ thickness is confined between the glass slides. The water film
contains dye-doped colloidal PS beads of $460\,\mathrm{nm}$ diameter. The motion of
the colloidal particles is monitored by widefield fluorescence microscopy, where
the fluorecence is excited at $532\,\mathrm{nm}$ wavelength by an expanded
laser beam ($\omega _{\mathrm{0,w}}\approx 20\,\mathrm{\mu m}$), collected
by an Olympus lens (100x/1.4) and imaged onto an Andor Ixon EMCCD camera in an inverted microscope. A
framerate of $100\,\mathrm{Hz}$ was used at a $2\times 2$ binning. An
additional focused laser beam ($\omega _{\mathrm{0,h}}\approx 1\,\mathrm{\mu m}$) also of $532\,\mathrm{nm}$ wavelength can be
steered in the sample plane with the help of an acousto-optic deflector (AOD) and is used for the plasmonic heating of the gold structure. The heating laser spot is driven in circles along the circumference of the gold structure at a rotation frequency $f=\omega/2\pi$. The data shown in Figures \ref{fig:fig5}, \ref{fig:fig7} and \ref{fig:fig8} were acquired on the same bead.

\section*{Acknowledgment}
This work was funded by the European Union and the Free State of Saxony. Also, financial support by the graduate school BuildMoNa and the Deutsche Forschungsgemeinschaft DFG (SFB TRR102) is acknowledged.

\printbibliography

\end{document}